\documentstyle[aps,pre,tighten,preprint]{revtex}
\begin{document}
\title{Spherical harmonic decomposition applied to spatial-temporal 
analysis of human high-density EEG
}
\author{Brett. M. Wingeier\cite{byline}}
\address{
Brain Sciences Institute, Swinburne University of Technology,\\
400 Burwood Road, Hawthorn, Victoria 3122, Australia.
}
\author{Paul. L. Nunez}
\address{Department of Biomedical Engineering, Boggs Center, 
Tulane University,\\
New Orleans, Louisiana, 70118.
}
\author{Richard. B. Silberstein}
\address{
Brain Sciences Institute, Swinburne University of Technology,\\
400 Burwood Road, Hawthorn, Victoria 3122, Australia.
}
\date{\today}
\maketitle
\begin{abstract}
We demonstrate an application of spherical harmonic 
decomposition to analysis of the human electroencephalogram (EEG). 
We implement two methods and discuss issues specific to analysis 
of hemispherical, irregularly sampled data. Performance of the 
methods and spatial sampling requirements are quantified using 
simulated data. The analysis is applied to experimental EEG data, 
confirming earlier reports of an approximate frequency-wavenumber 
relationship in some bands.
\end{abstract}

\narrowtext

\section{INTRODUCTION}
\label{sec:intro}
    The human electroencephalogram (EEG), as measured at the scalp, 
represents a superposition of electric fields resulting from 
post-synaptic potentials in neocortex, the thin (2 to 5 mm) surface 
layer of human brains. Several models of neocortical dynamics 
treat EEG as a mixed global/local phenomenon \cite{ref1,ref2,ref3},
and a better understanding of its spatial-temporal dynamics is necessary for 
evaluation and refinement of these models. Its temporal behavior 
has been studied at length, both by clinical observation \cite{ref4} 
and with such tools as power spectra \cite{ref5}, coherence \cite{ref6}, the 
Hilbert transform \cite{ref7}, and many others. However, until recently 
poor spatial resolution (due to minimal electrode sampling and 
under-use of head models) has limited spatial analysis of EEG 
\cite{ref1,ref2,ref8}.

    As a potential field on a near-hemispherical surface, EEG is 
amenable to analysis by spherical harmonic decomposition. In 
this paper, we apply two methods of decomposition (one 
described by Cadusch \cite{ref9} and one adapted from Misner \cite{ref10}) to 
131-channel EEG data. Using simulated data, we discuss issues 
and pitfalls relevant to such an analysis, specifically the effects 
of limited and irregular sampling density, integration over a 
hemisphere, and deviations from a spherical surface. From the 
experimental data, we then draw conclusions regarding the frequency-wavenumber 
relation of neocortical activity. 

\section{METHOD}
\label{sec:method}
We use the real spherical harmonics \cite{ref11}, defined on the sphere
$\Omega$ and described by the orthogonality integral
\begin{equation}
\langle Y_{lm} | Y_{l^{\prime}m^{\prime}} \rangle
=\int\limits_{0}^{2\pi }\int\limits_{0}^{\pi }Y_{lm}( \theta
,\phi) Y_{l^{\prime}m^{\prime}} ( \theta ,\phi ) \sin \theta
d\theta d\phi  =\delta _{l,l^{\prime}} \,\delta _{m,m^{\prime}}\label{eq1}.
\end{equation}
In theory, a potential field $\Phi ( \Omega )$ may be decomposed into 
spherical harmonic amplitudes $\Phi_{lm}$ defined by
\begin{equation}
\Phi _{lm} =\int\limits_{\Omega } Y_{lm} (\theta ,\phi )\Phi
(\theta ,\phi )d^{2} \Omega \label{eq2}.
\end{equation}
In the example of EEG and similar data we encounter three major 
and two minor issues.

\subsection{Sampling}

First, when attempting to decompose experimental data, we sample 
$\Phi ( \Omega )$ at specific locations $\Gamma$. Assuming near-regularly 
spaced electrodes, our maximum resolvable $l$ is determined 
by a spherical analog of the familiar Nyquist limit \cite{ref12} $f_{max} = 1
/ ( 2 \Delta T)$. With mean angular inter-electrode distance $\gamma$, 
we initially adopt a conservative limit of
\begin{equation}
l_{\max } =\left[ \frac{\pi }{2\gamma } \right]\label{eq3},
\end{equation} or $l_{max} = 6$ for our 131-channel electrode cap. Analog
pre-filtering to avoid spatial aliasing is not required here due to the
low-pass characteristics of the head volume conductor \cite{ref13}.

\subsection{Regularization}

With sampled data $F( \Gamma )$, the discretized version of the decomposition 
in Eq.\ \ref{eq2} is unstable in higher $l$-indices. (An apparently accurate 
reconstruction of the signal may be generated, with large artifacts 
in the higher spatial frequencies.) We must invoke constraint 
or regularization techniques to address this issue. Cadusch \textit{et 
al} \cite{ref9} approached the problem as a side issue of spherical spline 
interpolation. The estimate $\hat{\Phi } _{lm}$ is constrained by the spline
constraints, and the problem for a given sampling grid and $l_{max}$ is
reduced to a multiplication by a matrix of $ \mu_{lm}$  coefficients:
\begin{equation}
\hat{\Phi } _{lm} = \sum\limits_{x \in \Gamma }\mu _{lm} (x) F(x) \label{eq4}.
\end{equation}

Recently, Misner \cite{ref10} introduced a more complex method for
decomposition on a rectangular three-dimensional grid; that is, generalized 
to use $Y_{nlm}(r, \theta, \phi)$. We implement here the 
special case of sampling on a spherical surface, more relevant 
to EEG analysis, in which $r$ is constant. In Misner's method, 
a correction matrix of $G_{AB}$ accounts for discretization 
and limited sampling:
\begin{equation}
G_{AB} =\sum\limits_{x \in \Gamma }Y_{A} (x) Y_{B}
(x) w_{x}
\label{eq5},
\end{equation}
where A and B refer to index groups $(lm)$. Here we use the 
real harmonics and replace Misner's weight function $w_{x}$ with 
the effective area of each electrode. $G^{AB}$, the matrix 
inverse of the $G_{AB}$, is used to calculate ``adjoint 
spherical harmonics'' $Y^{A}$. Finally, a set of coefficients $R_{lm}$, 
analogous to Cadusch' $\mu_{lm}$, is generated and used 
as in Eq.\ \ref{eq4} to estimate $\hat{\Phi } _{lm} $.

\subsection{Hemispherical sampling}

Particularly relevant to EEG analysis is the error introduced 
by sampling over only half of the sphere. This sampling corresponds 
to only $\frac12$ of a spatial cycle of the $l = 1$, $m = 
0$ function, suggesting potential accuracy problems for functions 
involving $l = 0 \text{ and } 1$. It is also clear that the functions $Y_{lm}$ 
will no longer be orthonormal for $0 \leq \theta \leq \pi/2$ only; rather, 
we replace the $\delta$ in Eq.\ \ref{eq1} with an error quantity $\epsilon$:
\begin{equation}
\left\langle Y_{A} | Y_{B} \right\rangle
= \int\limits_{0}^{2\pi }\int\limits_{0}^{\frac{\pi }{2} }Y_{A}
\left( \theta ,\phi \right) Y_{B} \left( \theta ,\phi \right) \sin
\theta d\theta d\phi  =\epsilon _{AB} \label{eq6}.
\end{equation} In general, our hemispherical estimates
${\hat{\Phi ^ \prime}}_{A} $
will be related to the hypothetical full-sphere result by the 
matrix of $\epsilon_{AB}$'s:
\begin{equation}
{\hat{\Phi ^ \prime}}  =\bbox{\epsilon}\hat{\Phi } \label{eq7} 
\end{equation} and our ${\hat{\Phi ^ \prime}} _{A} $ become somewhat ambiguous
between certain sets of $(l,m)$. Although it may seem appropriate to invert
$\bbox{\epsilon}$ and calculate 
a more accurate result, the matrix is ill-conditioned ($R>10^{8}$, 
where the 2-norm condition number $R$ is the ratio of the largest 
singular value of $\bbox{\epsilon}$ to the smallest) and thus the inversion 
is problematic.

In addition, if we use the hemispherical region to calculate 
Misner's $G_{AB}$ in Eq. \ref{eq5}, the resulting matrix of $G_{AB}$ is
ill-conditioned ($R>10^{8}$) and thus the $G^{AB}$ cannot be reliably found. 
Rather, we created a mirrored set of electrodes $\Gamma ^ \prime$, calculated 
the matrix \textbf{R} of $R_{lm}$ for sample set
$(\Gamma \cup {\Gamma ^ \prime})$, and
discarded the antipodal rows of \textbf{R}\@.  Cadusch' spline 
method, while still subject to Eq. \ref{eq7}, is native to the hemispherical 
surface and requires no further manipulation.

\subsection{Coordinate orientation}

In many problems, the sphere has no preferred direction. The 
$m$-indices are usually collapsed \cite{ref14} to produce an angular power 
spectrum estimate $\hat{G} \left( l\right) $ as a function of wavenumber $l$:
\begin{equation}
\hat{G} (l) = \sum\limits_{m=-l}^{l} \left(\hat{\Phi ^ \prime } 
_{lm} \right) ^{2} \end{equation}
which is independent of coordinate system orientation. As well, 
we found the ``hemispherical error'' in $l$-spectrum 
to be independent of orientation. In some EEG studies, of course, 
the orientation of the underlying cerebral hemispheres may be 
relevant. In such cases, local spatial Fourier analysis \cite{ref1} should 
adequately complement our decomposition without the complication 
of distinguishing $m$-modes.

\subsection{Non-spherical media}

We assume that our medium $\Omega$ is a sphere, whereas biological 
data is often sampled on an irregular surface. The upper surface 
of the ``average'' human head \cite{ref15} may be represented 
as a hemiellipsoid with axes $a=10.52$ cm, $b=7.66$ cm, and $c=8.41$ 
cm, or alternatively 25\%, -9\%, and 0\% elongation from a perfect 
sphere. Although prolate spheroidal harmonics have been applied 
to biophysical field problems \cite{ref16,ref17,ref18}, the technique is often 
unwieldy. In comparison to error from $\epsilon_{AB}$, especially 
for low $l$, we assume the error due to approximating the ellipsoidal 
surface with a spherical surface is negligible. 

\section{APPLICATION TO SIMULATED DATA}

We generated evenly tessellated, hemispherical electrode maps 
of 74, 187, 282, and 559 electrodes, in addition to common experimental 
maps of 20, 64, and 131 \cite{ref8} electrodes. Five hundred potential 
maps were simulated for each electrode configuration. Each potential 
map was randomly generated with harmonics of degree $l = 6$, 
such that the $\Phi_{lm}$ varied with uniform distribution between 
0 and 1\@. Power spectrum estimates $\hat{G} \left( l\right) $ were then
calculated for each map, using both methods. Figure\ \protect\ref{corrunif}
shows Pearson's correlation coefficients $r_{l}$, calculated between
$G\left(l \right)$ 
and $\hat{G} \left( l\right) $ over the 500 trials for each electrode map.

We have noted that the error due to $\epsilon_{AB}$ causes power 
from one $(l,m)$-component to be misinterpreted as power in 
another, often of different $l$\@. Therefore, we might expect 
either method's performance to depend on the $l$-spectrum being 
analyzed. Using preliminary experimental data, we constructed 
an approximate power spectrum $G_{norm} (l)$ for average-referenced 
scalp EEG, peaked at $l=1$ and $l=2$, and decaying with $l^{-1}$ 
thereafter. Another five hundred potential maps were generated, 
with $\Phi_{lm}$ uniformly distributed between
\begin{equation}
0 < \Phi_{lm} < {G_{norm}(l) \over 2 l + 1}
\end{equation} to simulate a physiologically realistic distribution of
$l$-spectra. Power spectrum estimates $\hat{G} \left( l\right) $
 were calculated for each map using both methods. Results are 
shown in Fig.\ \protect\ref{corrreal}.

In general, results for the spline method --- though often quite 
accurate --- were dependent on the distribution of $l$-spectra 
being measured, exact electrode positions, and electrode numbers. 
Results for the adjoint harmonic method seemed more robust, even 
for sparse ($n = 64$) sampling, although accuracy was somewhat 
less in the higher harmonics.

In both methods, for $l = 6$ we observed minimal improvement 
for more than 131 electrodes. We thus believe that our 131-channel 
sampling is an appropriate tool for further study. Furthermore, 
given the limit in Eq.\ \ref{eq3} and the known volume-conductor attenuation 
of higher modes \cite{ref13}, we suggest that study of spatial frequencies 
higher than approximately $l = 8$ will be better 
served by intracranial EEG than by denser electrode maps.

In general, for low $l$ the adjoint harmonic method seemed 
more consistent. We examined typical 131-channel decompositions 
(Fig.\ \protect\ref{threemeths}) to investigate further. Both methods
accurately reproduced the potential maps ($r > 0.9$ for 131 channels). The
spline method, however, was slightly unstable for low $l$, and the 
erroneous negative 
$\hat{\Phi } _{lm} $
 are reflected in the power spectrum.

\section{REFINEMENTS AND ANALYSIS OF ERROR}

Any application of the spherical harmonic decomposition should 
take into account the estimated relative contribution of various 
error sources. Aside from measurement and experimental error, 
these may be divided into three categories: sampling error, orientation 
error, and hemispherical error.

\subsection{Sampling error}

Figures\ \protect\ref{corrunif} and\ \protect\ref{corrreal} indicate minimal
improvement for $l \leq 6$ with 
more than 131 electrodes. We can thus deduce that the Nyquist-like 
limit in Eq.\ \ref{eq3} is an appropriate guideline. When using coarser 
sampling, we expect some decrease in performance for higher $l$. Decreased 
accuracy for 20, 64, and 74 channels (seen in Figs.\ \protect\ref{corrunif}
 and\ \protect\ref{corrreal}, 
particularly for 20 channels) may be attributed to sampling error.

\subsection{Orientation error}

For a given $l$-spectrum, results will vary if power is randomly 
distributed across the $m$'s; that is, various $m$-components 
interact differently with our hemispherical sample grid. Five 
hundred random $l$-spectra, with realistic distribution of $G(l)$, were 
generated. Thirty 131-channel maps, randomly varying in $m$-power, 
were generated for each original $l$-spectrum. Figure\ \protect\ref{corrmult}
displays the resulting accuracy, again shown as correlation coefficient
$r_{l}$ between $G(l)$ and $\hat{G} \left( l\right) $ over the 100 trials.

Very little is gained in this simulation by decomposition of 
multiple epochs. As described in the following section, error 
due to $m$-distribution of power is largely swamped by hemispherical 
error. In practice, however, we must emphasize (in the presence 
of random measurement noise) the importance of averaging decompositions 
across many epochs. Orientation error will also become significant 
if our sampling grid is severely non-uniform.

\subsection{Hemispherical error}

In Sec.\ \ref{sec:method} above, we have discussed the hemispherical error
$\epsilon_{AB}$. Although it is impossible to improve our decomposition
results by inverting the matrix $\bbox{\epsilon}$, we may generate a
corresponding matrix for the power spectrum result and use it to estimate the 
contribution of hemispherical error.

Power in a single harmonic $Y_{lm}( \theta,\phi)$ is blurred 
by the hemispherical decomposition into surrounding harmonics. 
Using the 131-channel sampling map, we generated five hundred 
potential maps for each of $l=0\dots 6$, each with one unit power 
distributed randomly among the available $m$'s. By averaging 
over the five hundred resulting power spectra, for each $l$, 
we obtained an empirical ``averaged blurring matrix'' $\bbox{\rm{E}}$ 
for power spectra obtained from hemispherical decomposition. 
That is,
\begin{equation}
\hat{G} (l)\approx \bbox{\rm{E}} G ( l ) \label{eq10}.
\end{equation}
The typical $\bbox{\rm{E}}$ for both methods is a blurred identity matrix; 
that is, error in power spectra is largely between adjacent $l$. 
It is again tempting to invert $\bbox{\rm{E}}$, de-blur our spectra, and 
calculate a more accurate result, but although most $\bbox{\rm{E}}$ are 
invertable we found that for realistic spectra the benefit was 
marginal at best. Instead, $\bbox{\rm{E}}$ may be used to better understand 
the implications of hemispherical error.

We calculated correlation coefficients as in Fig.\ \protect\ref{corrmult},
between $\bbox{\rm{E}} G(l) \text{ and } \hat{G} \left( l\right) $
over the 500 trials of 30 epochs for 131 electrodes. The resulting 
higher correlations (although not applicable to a decomposition 
of real data) are plotted in Fig.\ \protect\ref{corrsph}. By comparison
with Figs.\ \protect\ref{corrreal}a and\ \protect\ref{corrmult}, the result
indicates the importance of hemispherical error. In particular, after
examination of typical $\bbox{\rm{E}}$ and $\bbox{\epsilon}$ 
matrices, we may interpret the decreased performance at low $l$ 
as blurring between adjacent wavenumbers. Furthermore, the increased 
effect, seen in Fig.\ \protect\ref{corrsph}, of averaging across various
$m$-distributions indicates that some abrupt changes in performance may be
attributed to sensitive interactions between $\bbox{\rm{E}}$-blurring and
random $m$-distribution.

Practically, the near-identity character of $\bbox{\rm{E}}$ is extremely 
useful. Hemispherical error manifests as blurring between adjacent $l$. 
Thus, we may expect composite measures such as the sum of power 
in $l$=0,1 to be substantially more accurate than individual 
estimates. Figure\ \protect\ref{corrcomposite} displays the accuracy of
$l$=0,1 and $l$=2\dots 6 adjoint harmonic power estimates (used below in
our experimental trials), for 500 epochs, realistic $l$-distribution, and
various sampling densities.

\section{APPLICATION TO EXPERIMENTAL DATA}

Nunez in 1974 \cite{ref19,ref20} and Shaw in 1991 \cite{ref21}, using Fourier
analysis along linear electrode EEG arrays, observed a relationship between 
increasing spatial and increasing temporal frequency in the 8-13 
Hz band, roughly consistent with simple wave dispersion relations. 
We attempted to duplicate this result in order to test the adjoint 
harmonic method under experimental conditions. We analyzed 131-channel 
EEG (resting, eyes closed) in five human subjects. Temporal Fourier 
coefficients were determined for 300 to 600 one-second epochs 
(depending on available data), and $l$-spectra averaged over 
these epochs.  

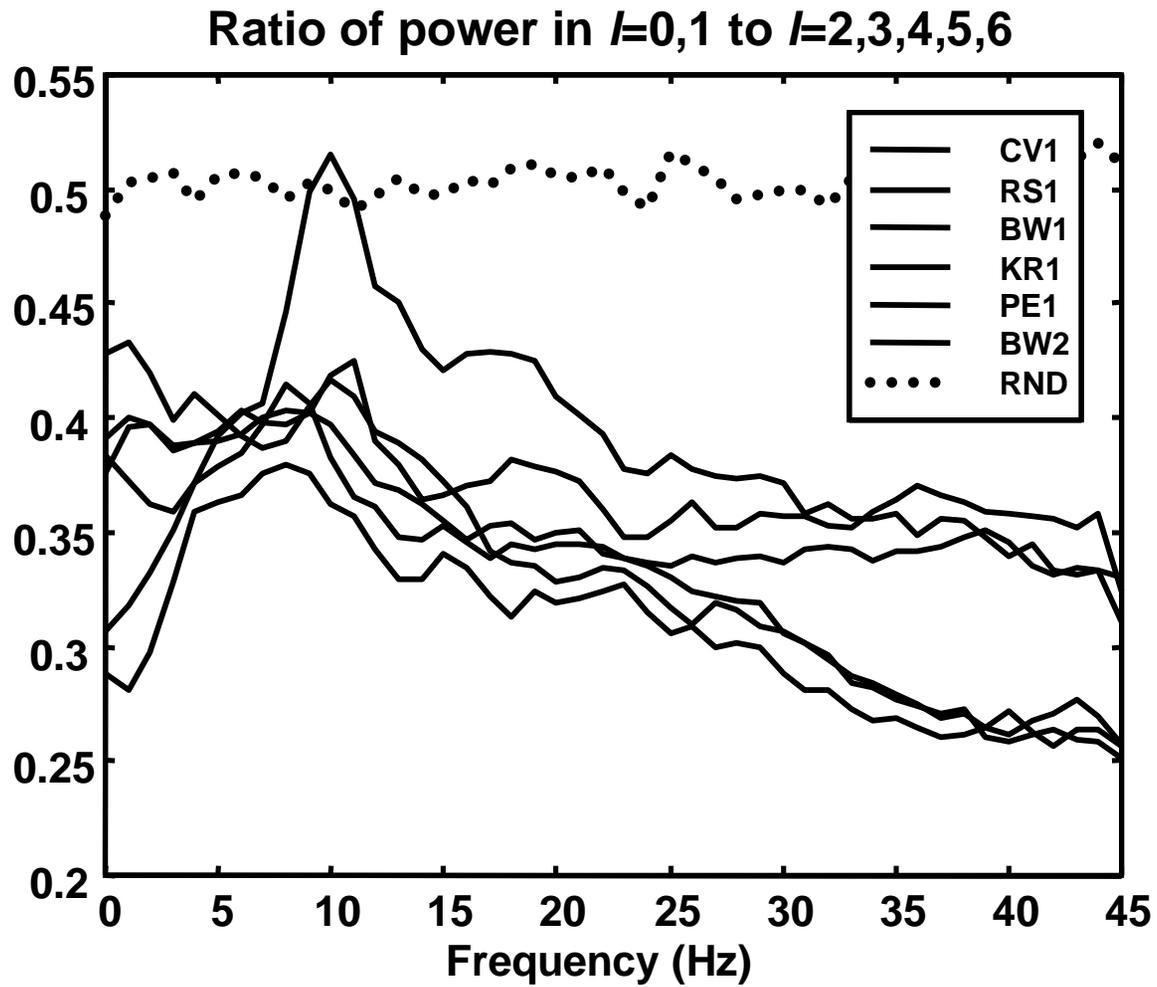
\begin{figure}
\caption{Five to ten minutes of resting, eyes-closed EEG were 
collected with 131 channels from each of six subjects (one duplicated). 
Complex temporal Fourier coefficients were calculated for one-second 
epochs and subjected to spherical harmonic spatial decomposition 
using the adjoint harmonic method. Resulting wavenumber spectra 
were averaged for each 1-Hz band over the 300 to 600 epochs. The 
ratio of power in $l$=0,1 to $l$=2,3,4,5,6 is plotted as a 
simple indicator of a bias toward higher spatial frequencies 
at higher temporal frequencies (greater than about 10 Hz). This 
result is qualitatively consistent with the postulated existence 
of an approximate EEG dispersion relation, perhaps with alpha 
rhythm (8--13 Hz) representing the fundamental and lower overtones.
A surrogate signal (dotted line), composed of random EEG-like noise
and subjected to the same analysis, showed no such relation.}
\label{experiment}
\end{figure}

Results are summarized in Fig.\ \protect\ref{experiment} as the ratio of
power in low ($l=0,1$) to power in high ($l=2,3,4,5,6$) spatial frequencies. 
Above approximately $f = 8\text{ Hz}$, with increasing $f$ we observed 
a general trend towards power in higher $l$. We also observed 
high-wavenumber power in the delta band ($f \leq 3\text{ Hz}$). The 
alpha band (c. 8--13 Hz) was characterized by the highest power 
in low spatial frequencies.

In order to rule out methodological artifact, we generated and analyzed 300
seconds of simulated EEG using 3602 uncorrelated sources, each generating 
1/f noise through a head volume conductor model as described 
in \cite{ref13}.  As expected, the EEG-like noise (labeled RND in the figure)
showed no relation between spatial and temporal frequencies.

\section{CONFIDENCE INTERVAL ESTIMATION}

Estimates of the temporal power spectrum are known to vary in 
chi-square distribution \cite{ref12}, assuming normally distributed estimates 
of the underlying Fourier coefficients. Error distribution for 
spatial spectrum estimates, on the other hand, is complicated 
by the dependence of hemispherical and orientation error on the 
entire $l$-spectrum. For composite measures of both spatial 
and temporal spectra, such as shown in Fig.\ \protect\ref{experiment}, the
situation becomes even more problematic. We propose an empirical test for 
estimation of such confidence levels, analogous to the randomization 
tests commonly applied in nonparametric statistical analysis 
\cite{ref22}.

Let
\begin{equation}
A = {{G_{01} }\over{G_{2...6} }},
\end{equation} where $G_{01}$ is the total power in harmonics $l$=0 and
$l$=1, and $G_{2\dots 6}$ is the total power in harmonics $l$=2 through
$l$=6. Let $\hat{G} _{01} $, $\hat{G} _{2...6} $, and $\hat{A} $
represent estimates of the same. Above, we calculated 
$\hat{A} _{f} $ for various temporal frequencies. Here, we calculate an
approximate 95\% confidence interval for single-epoch estimates of the
actual $A_{f}$. The confidence interval will apply only to the spatial
spectrum composite measure, neglecting error (or nonstationarity) in temporal 
frequency spectra, which for many applications may be as important. 
Note, though, that for 300 epochs the normalized standard error 
of a temporal power spectrum estimate is less than 6\%.

To determine an empirical confidence interval, we would typically 
examine the distribution of random re-samples. In this application, 
we created many random $l$-spectra from an estimated distribution 
of $l$-power, simulated many decompositions, and examined the 
resulting distribution as follows.

Since hemispherical error is dependent on $l$-spectrum, the 
result will be influenced by the distribution of the random spectra. 
Srinivasan \textit{et al} \cite{ref13} analytically estimated the spatial 
frequency domain transfer function for volume-conduction blurring 
of scalp potential as proportional to $(2 l + 1)^{-1}$. This ``spatial 
smearing'' is due mainly to the poorly conducting skull 
and physical separation between cortical current sources and 
scalp electrodes. In our calculation, we assumed that underlying $l$-spectra 
vary in uniform distribution in proportion to $( 2 l + 1)^{-1}$, 
and that with average-referenced data the contribution of $l$=0 
is negligible. A large number (20,000) of $l$-spectra were 
generated, randomly selecting for each $l$-bin a value from 
the appropriate distribution. The decomposition was performed, 
and the composite measure $A$ calculated, for each randomized spectrum.

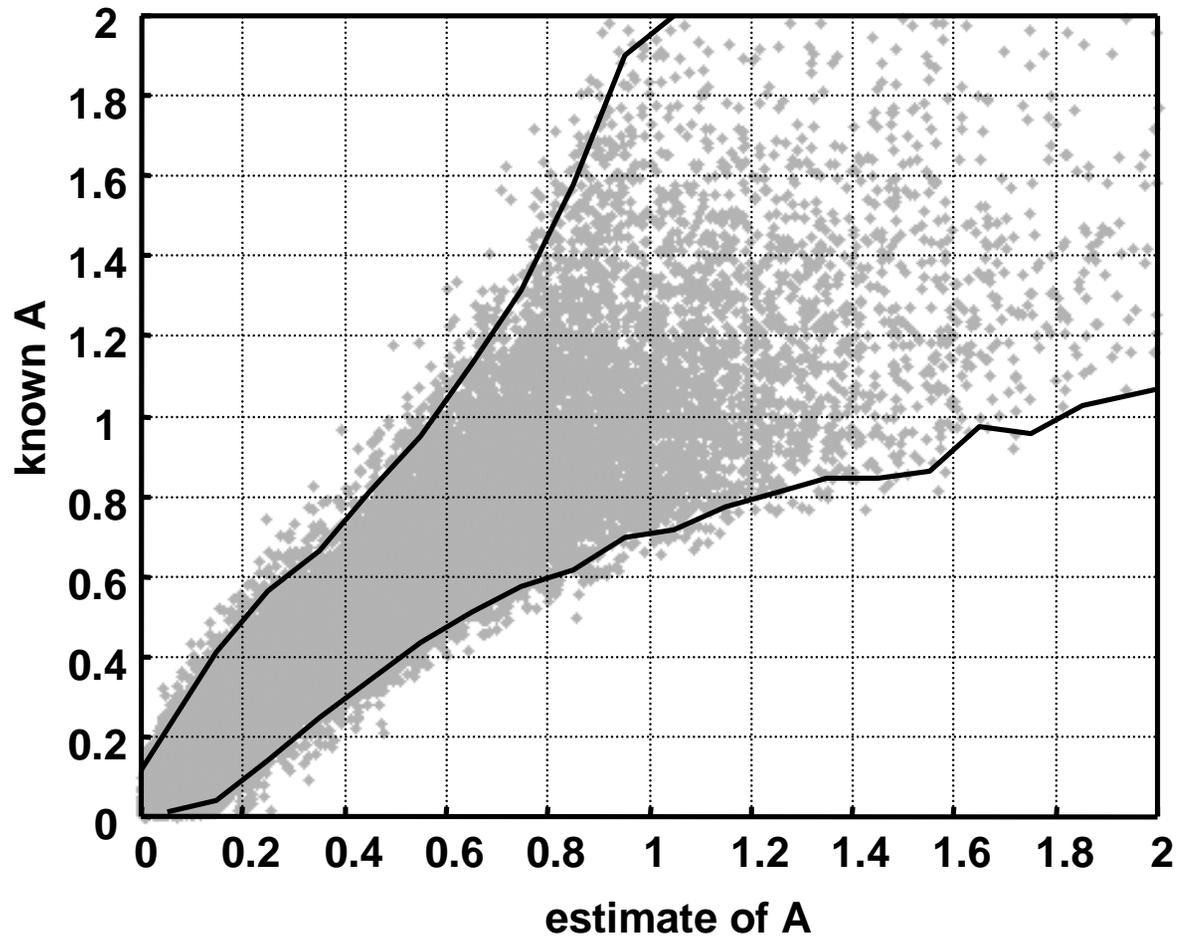
\begin{figure}
\caption{95\% confidence intervals for single-epoch estimates 
of the power ratio $A = {G_{l=0,1}} / {G_{l=2 \dots 6}}$. 
Twenty thousand potential maps were generated from known, random, 
realistically distributed (based on genuine EEG data) $l$-spectra, 
and decomposed using 131 channels and the adjoint harmonic method. 
Here, known $A$ are plotted against the resulting estimated $A$. 
Solid lines indicate the empirical 95\% confidence interval for 
a given estimate of $A$. Multiple-epoch estimates will result 
in much smaller intervals, depending on the variation in $l$-spectra 
being decomposed.}
\label{confidence}
\end{figure}

By examining the distribution of known surrogate 
$A_{rand} $ which produce a certain estimate 
$\hat{A} _{rand} $
, we can estimate an empirical confidence interval for our spectral 
estimate. In Fig.\ \protect\ref{confidence}, we show the scatter plot of 
$A_{rand} $
 against 
$\hat{A} _{rand} $
 with 95\% confidence intervals. For a given estimate 
$\hat{A} $
 and the assumptions discussed above, 95\% of the time, the actual $A$ 
will fall between the two lines shown.

A similar procedure may be used to calculate confidence intervals 
for other measures, whether the actual $G_{l}$ or other composite 
measures. Careful judgment must be applied when estimating confidence 
intervals for multiple-epoch measures such as shown in
Fig.\ \protect\ref{corrcomposite}. As demonstrated earlier in this paper,
variation in the $m$-component of an $l,m$-spectrum only allows us
to ``average out'' the minimal orientation error. Variation in hemispherical
error (dependent on $l$-spectrum), without gross violation of the 
stationarity assumption, is necessary for the average of estimates 
$\hat{A} $
 over multiple epochs to converge to $A$.

\section{DISCUSSION}

Our simulations provide a firm basis for application of spherical 
harmonic decomposition to irregularly sampled, hemispherical 
data such as EEG. Our hemispherical modification of Misner's 
adjoint harmonic method \cite{ref10} proved most consistent. However, 
for physiological data of known power distribution, the spline 
method \cite{ref9} is complementary and may be slightly more accurate 
with high-density sampling. It seems that, within the conservative 
band-limit of equation \cite{ref3} and the known spatial filter properties 
of the head \cite{ref13}, decomposition accuracy will not be materially 
improved by more than 131 electrodes for scalp EEG. We suggest 
that confidence intervals for such decompositions, or for
decomposition-derived measures, be determined empirically using randomized
data. Furthermore, while single-decomposition errors are relatively large,
with multiple epochs the experimental accuracy may be increased
substantially. For this averaging to be both valid and effective, we must
assume a quasi-stationary wavenumber spectrum across our epochs, but 
with sufficient random variation in hemispherical error for our 
estimates to converge upon the mean. In addition, especially 
in EEG applications, we must remain aware of the limitations 
inherent in collapse across $m$'s (we assume the orientation 
of the underlying cerebral hemispheres is irrelevant) and the 
use of spherical harmonics on a hemispheroidal surface.

The dynamical properties of human EEG rhythms are quite complicated, 
varying substantially between individuals and brain states. Furthermore, 
physiologically-based theoretical models point to substantial 
nonlinear effects and interactions across spatial scales
\cite{ref2,ref23,ref24,ref25,ref26}. Despite all the obvious complications,
results from the spherical harmonic decomposition of experimental EEG agreed
qualitatively with crude linear electrode array results \cite{ref1,ref21}.
These results were seen in all subjects and are consistent with a mixed
global/local model of cortical dynamics, in which lower global mode
oscillations produce alpha rhythm, superimposed on local (spatially
uncorrelated) activity in various frequency bands \cite{ref2}. Further study
of spatiotemporal EEG dynamics, using spherical harmonic decomposition,
should shed more light on these issues.

\acknowledgements

This work was supported by the Australian Research 
Council grant \#A10019013 and by the U.S. National Science Foundation 
(B.M.W.).

\begin{figure}
\caption{Correlations between actual and estimated $l$-power 
for uniformly distributed random spectra. Five hundred potential 
maps were generated from known wavenumber spectra, with random 
power in each $l$-component, uniformly distributed between 
0 and 1. For each of seven electrode densities, wavenumber spectra 
were estimated by spherical harmonic decomposition of the 500 
sampled maps. Correlations between actual power and estimated 
power were calculated over the 500 trials for each $l$-component. 
Shown here for (a) adjoint harmonic and (b) spline methods, these 
correlations are a measure of the quality of a single decomposed 
power spectrum.}
\label{corrunif}
\end{figure}

\begin{figure}
\caption{Correlations between actual and estimated \textit{l}-power 
for more realistically distributed random spectra based on genuine 
EEG data. As in Fig.\ \protect\ref{corrunif}, but in original spectra random
power in each $l$-component is uniformly distributed between 0 and 
$(2 l + 1)^{-1}$.}
\label{corrreal}
\end{figure}

\begin{figure}
\caption{Topography (left column), $l,m$-spectra (center column), 
and $l$-power (right column) for a typical 131-channel spherical 
harmonic decomposition. The original map is shown in (a). The 
adjoint harmonic method (b) reconstructs topography and gives 
an approximation of $l$-spectrum. Although the spline method 
(c) also reconstructs potential topography, we observe irregularities 
in the lower $l$ amplitude estimates that contribute to decreased 
performance for these wavenumbers, and a less accurate $l$-spectrum 
estimate.}
\label{threemeths}
\end{figure}

\begin{figure}
\caption{Correlations between actual and estimated $l$-power 
for multiple-epoch, adjoint-harmonic estimates of the same $l$-spectrum, 
with epochs varying only in $m$-component. For a reasonably 
isotropic and dense sample array, such as the 131-channel EEG 
grid used here, there is little orientation error and thus little 
improvement in results.}
\label{corrmult}
\end{figure}

\begin{figure}
\caption{Sampling a full sphere with 262 channels and the adjoint 
harmonic method, correlations between actual and estimated $l$-power 
are shown for multiple-epoch (varying only in $m$-component) 
estimates of the same $l$-spectrum. By sampling over the full 
sphere, we eliminate hemispherical errors illustrated in
Fig.\ \protect\ref{corrmult}. Remaining errors are due to orientation (note
improvement with multiple epochs) and imperfect sampling.}
\label{corrsph}
\end{figure}

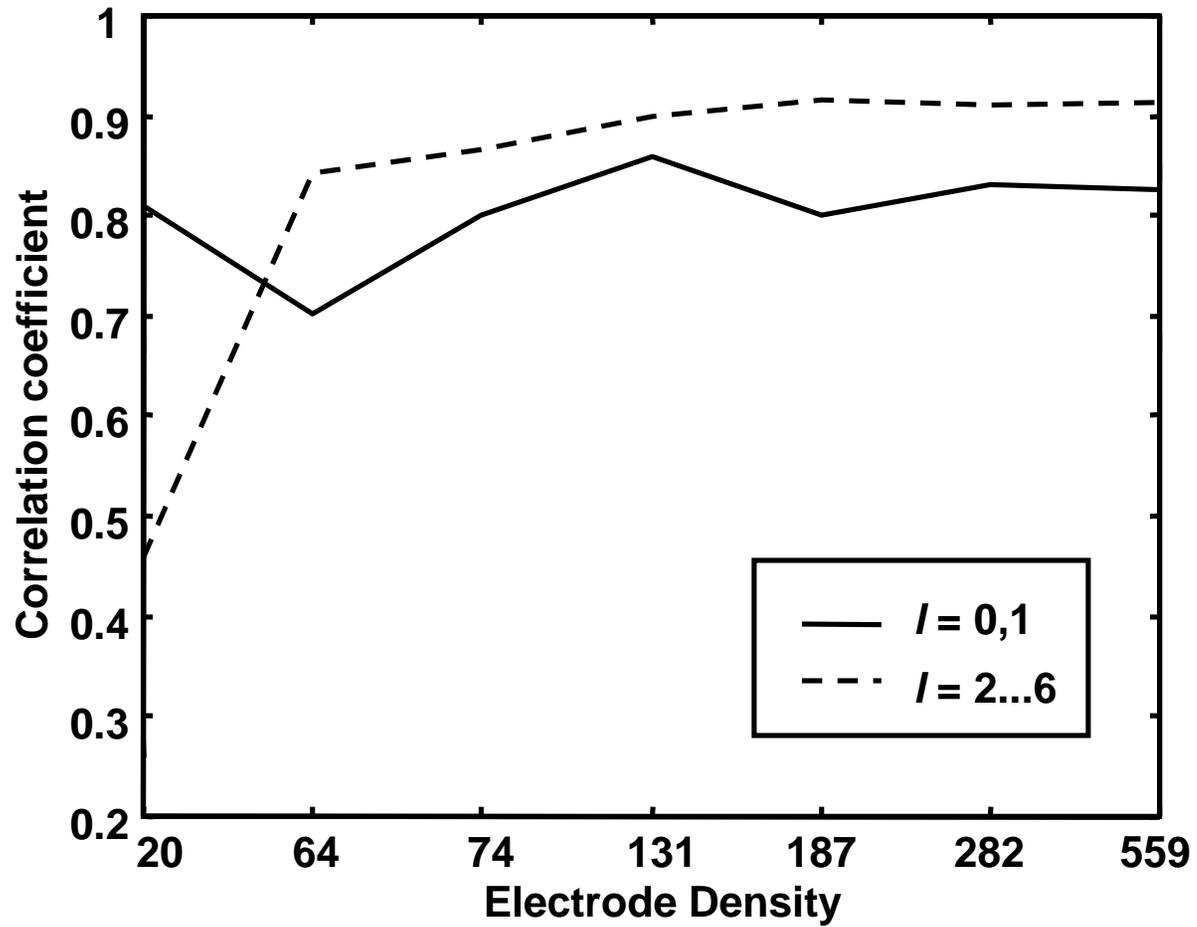
\begin{figure}
\caption{Correlation coefficients, using the adjoint harmonic 
method, obtained by comparisons of estimated to actual summed 
power measures. The solid line represents power in $l$=0 and $l$=1 
modes, and the broken line represents power summed over modes $l$=2 
through $l$=6. Increased accuracy (as compared to part A of
Fig.\ \protect\ref{corrreal}) is because most hemispherical error manifests
as blurring between power in adjacent $l$'s.}
\label{corrcomposite}
\end{figure}

\end{document}